\documentclass[a4paper]{panl}
\usepackage{cite}
\usepackage{wrapfig}
\usepackage{graphicx}
\usepackage{amssymb}
\usepackage{amsfonts}
\usepackage{amsmath}
\usepackage{longtable}
\usepackage{rotating}
\usepackage{lscape}
\usepackage{epsfig}
\usepackage{multirow}
\usepackage[colorlinks]{hyperref}

\originalTeX
\begin{document}

\title{Testing General Relativity with black hole X-ray data \\ Проверка общей теории относительности с помощью рентгеновских данных черных дыр}
\maketitle
\authors{Cosimo~Bambi\footnote{E-mail: bambi@fudan.edu.cn}}
\setcounter{footnote}{0}
\authors{Косимо Бамби\footnote{E-mail: bambi@fudan.edu.cn (русский вариант)}}
\from{Center for Field Theory and Particle Physics and Department of Physics,\\Fudan University, Shanghai 200438, China}
\from{Центр теории поля и физики элементарных частиц и кафедра физики,\\Фуданьский университет, Шанхай 200438, Китай}
\from{School of Natural Sciences and Humanities,\\New Uzbekistan University, Tashkent 100007, Uzbekistan}
\from{Школа естественных и гуманитарных наук,\\Университет Новый Узбекистан, Ташкент 100007, Узбекистан}

\begin{abstract}
The theory of General Relativity has successfully passed a large number of observational tests without requiring any adjustment from its original version proposed by Einstein in 1915. The past 8~years have seen significant advancements in the study of the strong-field regime, which can now be tested with gravitational waves, X-ray data, and black hole imaging. This is a compact and pedagogical review on the state-of-the-art of the tests of General Relativity with black hole X-ray data.

\vspace{0.2cm}

Общая теория относительности успешно прошла большое количество наблюдательных проверок, не требуя каких-либо корректировок по сравнению с ее первоначальной версией, предложенной Эйнштейном в 1915 году. За последние 8~ лет были достигнуты значительные успехи в изучении режима сильного поля, который теперь можно проверить с помощью гравитационных волн, рентгеновских данных и изображений черных дыр. Это компактный и педагогический обзор современного состояния тестов общей теории относительности с рентгеновскими данными черных дыр.
\end{abstract}
\vspace*{6pt}

\noindent

\section*{Introduction}\label{sec:intro}

The theory of General Relativity is one of the pillars of modern physics. The theory was proposed by Albert Einstein at the end of 1915. In more than 100~years, it has passed a large number of observational tests without requiring any modification from its original version. General Relativity has been extensively tested with Solar System experiments and observations of binary pulsars, which can mainly probe the so-called {\it weak field regime}~\cite{Will:2014kxa}. In the past 30~years, significant efforts have been devoted to test General Relativity on {\it large scales} with cosmological tests in order to investigate the problems of dark matter and dark energy~\cite{Ferreira:2019xrr}. More recently, thanks to a new generation of observational facilities, we have started testing General Relativity in the so-called {\it strong field regime} with black holes~\cite{Bambi:2015kza,Yagi:2016jml}. Today we have gravitational wave data of the coalescence of compact binary systems from the LIGO-Virgo-KAGRA Collaboration, images of the supermassive black holes M87$^*$ and SgrA$^*$ from the Event Horizon Telescope experiment, and X-ray observations of black holes from a number of X-ray missions. Here I will summarize the state-of-the-art of the tests of General Relativity with black hole X-ray data, discussing recent progress and future developments. More details can be found in the recent review in~\cite{Bambi:2022dtw} and references therein.

\section*{Disk-Corona Model}\label{sec:dc}

\begin{figure}[t]
\begin{center}
\includegraphics[width=0.95\linewidth]{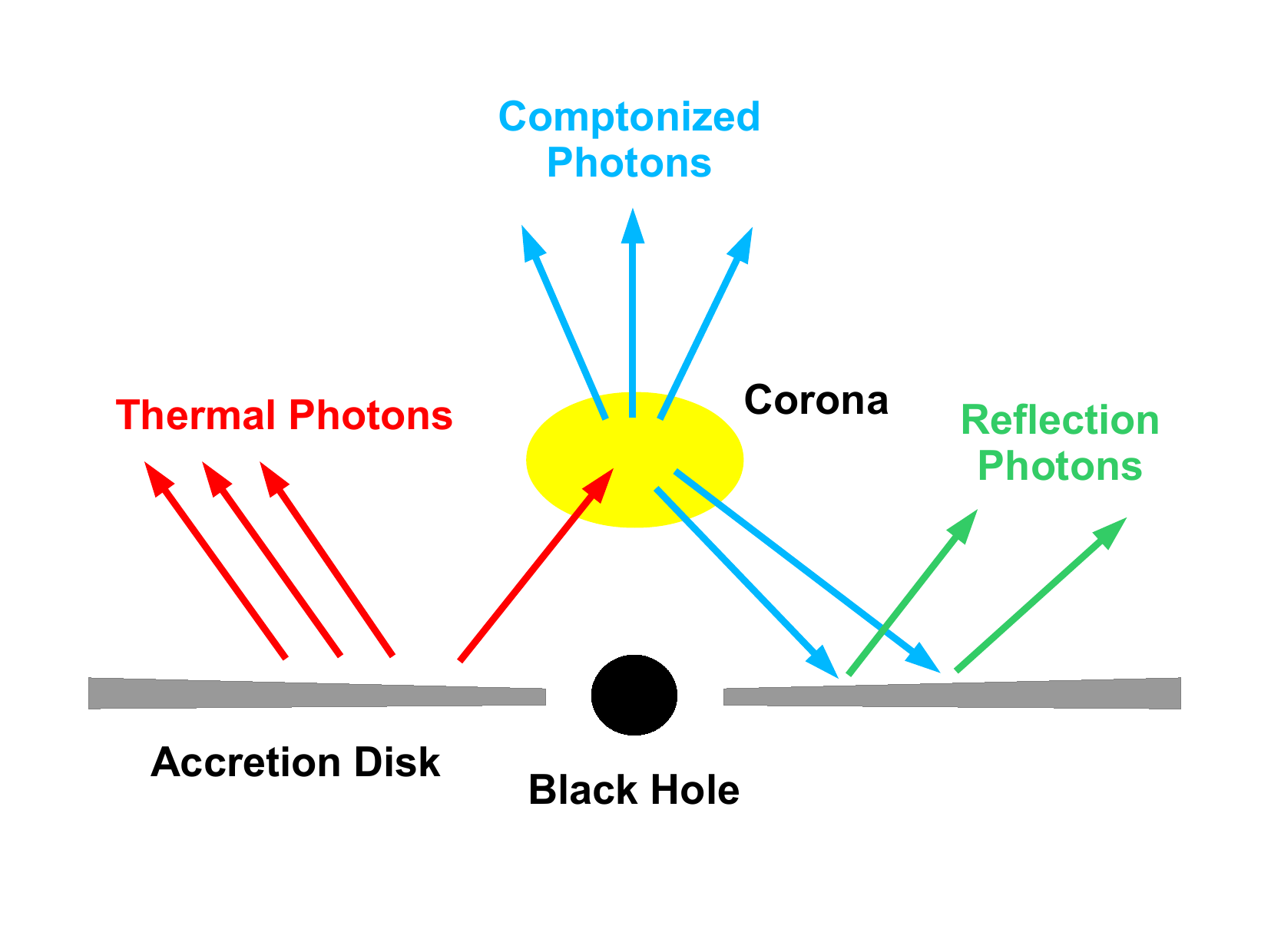}
\end{center}
\vspace{-1.8cm}
\caption{Disk-corona system. Figure from Ref.~\cite{Bambi:2021chr} under the terms of the Creative Commons Attribution 4.0 International License. \label{f-corona}}
\end{figure}

Fig.~\ref{f-corona} shows the system that we want to use for our X-ray tests of General Relativity. It is normally referred to as the disk-corona system~\cite{Bambi:2020jpe}. The central black hole can either be a stellar-mass black hole in an X-ray binary or a supermassive black hole in an active galactic nucleus. The key-point is that the black hole is accreting from a geometrically thin and optically thick accretion disk. The thermal spectrum of the accretion disk (red arrows in Fig.~\ref{f-corona}) is normally peaked in the soft X-ray band (0.1-10~keV) in the case of stellar-mass black holes and in the UV band (1-100~eV) for the supermassive ones. Thermal photons from the disk can inverse Compton scatter off free electrons in the ``corona'', which is some hotter plasma ($\sim$100~keV) near the black hole and the inner part of the accretion disk. The exact nature and geometry of the corona is not yet well understood: it may be the base of the jet, the hot atmosphere above the accretion disk, the material in the plunging region between the disk and the black hole, etc. A fraction of the Comptonized photons (blue arrows in Fig.~\ref{f-corona}) can illuminate the disk: Compton scattering and absorption followed by fluorescent emission generate the reflection spectrum (green arrows in Fig.~\ref{f-corona}). The reflection spectrum in the rest-frame of the material of the disk is characterized by narrow fluorescent emission lines in the soft X-ray band and a Compton hump with a peak at 20-30~keV. The reflection spectrum of the whole disk detected by a distant observer is blurred because of relativistic effects (Doppler boosting and gravitational redshift)~\cite{Bambi:2017khi}.

\section*{Testing Gravity with Black Hole X-ray Data}\label{sec:tests}

In 4-dimensional General Relativity, uncharged black holes are described by the Kerr solution. The spacetime geometry around astrophysical black holes formed from the complete gravitational collapse of stars or clouds is thought to be described well by the Kerr solution and deviations induced by nearby stars, accretion disks, possible non-vanishing electric charges, etc. are normally completely negligible~\cite{Bambi:2017khi}.

So far, most of the efforts to test General Relativity with black hole X-ray data have been devoted to check the so-called {\it Kerr black hole hypothesis}, namely if the spacetime around astrophysical black holes is described by the Kerr solution of General Relativity. However, other tests are also possible, including tests of the Weak Equivalence Principle or tests of the non-gravitational physics in strong gravitational fields (see Ref.~\cite{Bambi:2022dtw} for more details).

As for many other theories, there are two different strategies to test General Relativity and they are usually referred to as top-down and bottom-up, respectively. The {\it top-down} (or theory-specific) approach is the most logical one: we want to test General Relativity against another theory of gravity. In such a case, we can construct two models, one for General Relativity and one for the other theory of gravity. We fit the observational data with the two models and we check which of them can explain the data better and if we can rule out the other one. This strategy requires to be able to calculate the predictions of both General Relativity and the other theory of gravity, which is often not so easy. In the {\it bottom-up} (or agnostic) approach, we parametrize possible deviations from the predictions of General Relativity and we try to measure these extra parameters to check {\it a posteriori} whether observations are consistent with the predictions of General Relativity. An example of bottom-up approach is the Parametrized Post-Newtonian (PPN) formalism, which has been extensively used to test General Relativity in the Solar System~\cite{Will:2014kxa}.

In the past years, we have developed two models to test General Relativity with black hole X-ray data: {\tt relxill\_nk}~\cite{Bambi:2016sac,Abdikamalov:2019yrr} and {\tt nkbb}~\cite{Zhou:2019fcg}. Both models are public on GitHub\footnote{\href{https://github.com/ABHModels}{https://github.com/ABHModels}.} and can be used with the X-ray package XSPEC~\cite{xspec}. {\tt relxill\_nk} is a model to test General Relativity from the analysis of the relativistically blurred reflection features in the X-ray spectra of accreting black holes (X-ray reflection spectroscopy). {\tt nkbb} is a model to test General Relativity from the analysis of the thermal component of black hole accretion disks (continuum-fitting method). More details on these models can be found in the original publications describing the models and in Ref.~\cite{Bambi:2022dtw}. In the next sections, I summarize the main results.

\section*{Results: Agnostic Tests}\label{sec:results1}

The Johannsen metric~\cite{Johannsen:2013szh} is a parametric black hole spacetime specifically designed to test the Kerr hypothesis from black hole electromagnetic observations. It is characterized by four infinite sets of deformation parameters: $\{ \alpha_{1k} \}$, $\{ \alpha_{2k} \}$, $\{ \alpha_{5k} \}$, and $\{ \epsilon_{k} \}$. When all the deformation parameters vanish, we exactly recover the Kerr solution of General Relativity. The Johannsen metric is not a solution of any gravity theory, but it is supposed to be able to mimic a large number of black hole spacetimes for suitable choices of the values of its deformation parameters.

\begin{figure}[t]
\begin{center}
\includegraphics[width=0.95\linewidth]{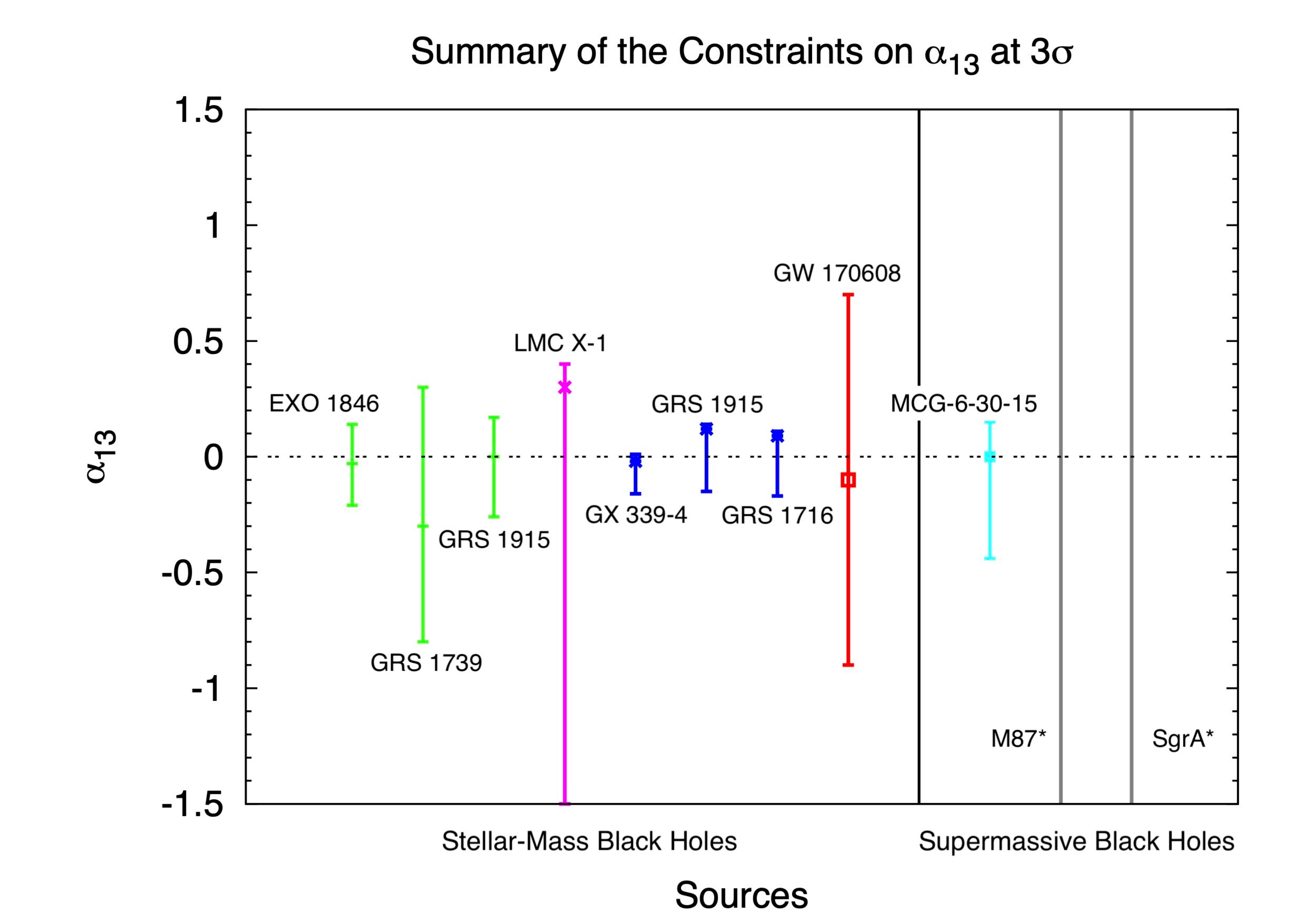}
\end{center}
\vspace{-0.6cm}
\caption{Summary of current 3-$\sigma$ constraints on the Johannsen deformation parameter $\alpha_{13}$ (the dotted horizontal line for $\alpha_{13} = 0$ corresponds to the Kerr solution). {\it Stellar-mass black holes}: constraints from X-ray reflection spectroscopy (green), continuum-fitting method (magenta), combination of X-ray reflection spectroscopy and continuum-fitting method (blue), and gravitational waves (red). {\it Supermassive black holes}: constraints from X-ray reflection spectroscopy (cyan) and black hole imaging (gray). See the text for more details. Figure from Ref.~\cite{cb22}. \label{f-summary}}
\end{figure}

For simplicity, most tests of the Kerr black hole hypothesis with the Johannsen metric have considered black hole metrics with a possible non-vanishing $\alpha_{13}$ and have assumed that all other deformation parameters vanish. Fig.~\ref{f-summary} summarizes current constraints on $\alpha_{13}$ from X-ray, gravitational waves, and black hole imaging. The dotted horizontal line at $\alpha_{13} = 0$ corresponds to the Kerr solution of General Relativity and current tests are all consistent with a vanishing $\alpha_{13}$. All error bars are at 3-$\sigma$, corresponding to the 99.7\% confidence level limits. In Fig.~\ref{f-summary}, we consider separately tests with stellar-mass and supermassive black holes because these object classes can potentially test two different curvature regimes.

In the case of stellar-mass black holes, we can constrain $\alpha_{13}$ from X-ray and gravitational wave data. The constraints in green are the most precise and accurate constraints on $\alpha_{13}$ from the analysis of black hole reflection spectra with {\tt relxill\_nk}~\cite{Tripathi:2020yts}. There are other constraints in the literature, but they are less precise and/or accurate and therefore are not reported in the figure. The constraint in magenta is from the continuum-fitting method with {\tt nkbb}~\cite{Tripathi:2020qco}: the 3-$\sigma$ error bar extends well beyond the $y$-axis range of the plot and the constraint is weak because of a strong degeneracy between the estimate of $\alpha_{13}$ and the black hole spin parameter. The constraints in blue are the results from some special sources in which we can constrain $\alpha_{13}$ by using the two techniques together, either because we have spectra with simultaneous strong blurred reflection features and prominent thermal components (this is the case of GX~339--4~\cite{Tripathi:2020dni} and GS~1716~\cite{Zhang:2021ymo}) or because we can combine the analysis of spectra with strong blurred reflection features with the analysis of spectra with prominent thermal components (as in the case of GRS~1915~\cite{Tripathi:2021rqs}). The constraint in red is the most stringent constraint on $\alpha_{13}$ that we can currently infer from gravitational wave data of the coalescence of two stellar-mass black holes~\cite{Shashank:2021giy}. More details on these constraints, the spectral analysis to infer them, and the systematic uncertainties can be found in the original publications~\cite{Tripathi:2020yts,Tripathi:2020qco,Tripathi:2020dni,Zhang:2021ymo,Tripathi:2021rqs,Shashank:2021giy}.

In the case of supermassive black holes, we can constrain $\alpha_{13}$ from X-ray reflection spectroscopy and black hole imaging. The continuum-fitting method cannot be used in this case because the thermal component of the accretion disks of supermassive black holes is peaked in the UV, where dust absorption strongly limits the possibility of accurate measurements. Current ground-based gravitational wave observatories are sensitive to gravitational wave frequencies ranging from a few Hz to a few kHz, and therefore we cannot observe gravitational waves from systems with supermassive black holes, since the frequency of the emitted gravitational waves is too low. The constraint in cyan is the most precise and accurate constraints on $\alpha_{13}$ from the analysis of black hole reflection spectra with {\tt relxill\_nk}~\cite{Tripathi:2018lhx}. In general, stellar-mass black hole data can provide stronger constraints on possible deviations from the Kerr geometry than observations of supermassive black holes. However, MCG--06--30--15 is quite a special source: it is bright, the spectrum can present a very broadened iron line, and we have high-quality data from simultaneous observations of \textsl{NuSTAR} and \textsl{XMM-Newton}. The constraint on $\alpha_{13}$ from MCG--06--30--15 is eventually comparable to the best constraints on $\alpha_{13}$ from stellar-mass black holes in X-ray binaries, but normally the constraints from supermassive black holes are less precise and accurate. The constraints in gray are from the images of the supermassive black holes in M87$^*$~\cite{EventHorizonTelescope:2020qrl} and SgrA$^*$~\cite{EventHorizonTelescope:2022xqj} from the Event Horizon Telescope experiment. The two 3-$\sigma$ error bars extend well beyond the $y$-axis range of the plot. These constraints are weak because they are currently limited by the poor angular resolution of the images. In the future, we can expect to have one of the stations on a satellite orbiting the Earth: with such a distant station, the Event Horizon Telescope experiment could improve its angular resolution by an order of magnitude and reach the precisions of current X-ray and gravitational wave tests of the Kerr metric.

\section*{Results: Tests of Specific Gravity Theories}\label{sec:results2}

The agnostic X-ray tests of the Kerr hypothesis with {\tt relxill\_nk} and {\tt nkbb} reviewed in the previous section can be repeated to test specific gravity theories in which uncharged black holes are not described by the Kerr solution. It is necessary to know the full rotating black hole solution of the theory of gravity that we want to test, implement such a metric in {\tt relxill\_nk} and/or {\tt nkbb}, and analyze high-quality black hole spectra. In the past years, this has been done for a few theories of gravity and can be potentially done for more, as long as we know their rotating black hole solutions.

Constraints have been obtained for Einstein-Maxwell-Dilaton-Axion Gravity~\cite{Tripathi:2021rwb}, a class of conformal gravity models~\cite{Zhou:2018bxk,Zhou:2019hqk}, and Asymptotically Safe Quantum Gravity~\cite{Zhou:2020eth}. These constraints from X-ray data are the most stringent constraints that can be found today in the literature for these models. The interested reader can refer to the original publications for more details~\cite{Tripathi:2021rwb,Zhou:2018bxk,Zhou:2019hqk,Zhou:2020eth}.

\section*{Concluding Remarks}\label{sec:conc}

As it was discussed in the previous sections and it is evident from Fig.~\ref{f-summary}, X-ray tests of General Relativity are currently very competitive with respect to other techniques and can provide the most {\it precise} measurements of possible deviations from the Kerr metric. The question is: are these measurements also {\it accurate}?

The actual accuracy of a measurement is always a crucial issue in astrophysics. My (somewhat personal) answer is that current X-ray tests of General Relativity can be both precise and accurate, but it is necessary to select carefully the right sources and observations. Indeed, while {\tt relxill\_nk} and {\tt nkbb} represent the state-of-the-art in the analysis of reflection features and thermal spectra for testing General Relativity, they are still relatively simple models with a number of simplifications. This means that these models can approximate better the spectra of some sources and worse the spectra of other sources. It is thus crucial to select the sources and observations suitable to obtain measurements that are both precise and accurate.

Relativistic signatures in X-ray spectra are stronger if the reflection component is strong and mainly generated in the very strong gravity region as close as possible the black hole, which, in turn, requires that the inner edge of the accretion disk is as close as possible to the black hole event horizon and the corona is compact and as close as possible to the black hole. It is also crucial to select sources with geometrically thin and optically thick accretion disks: theoretical and observational studies have shown that the simple Novikov-Thorne model with an infinitesimally thin disk works well as long as the disk is geometrically thin~\cite{Abdikamalov:2020oci,Shashank:2022xyh}, while we can easily obtain incorrect measurements of the properties of the system if the accretion disk of the source is not geometrically thin~\cite{Riaz:2019bkv,Riaz:2019kat}. A detailed discussion on the selection of spectra suitable for precise tests of General Relativity and on the systematic uncertainties in current X-ray reflection spectroscopy measurements can be found in Ref.~\cite{Bambi:2022dtw}.

While accurate tests of General Relativity are possible with present X-ray observatories as long as we select carefully the spectra to analyze, current reflection models rely on a number of simplifications. The next generation of X-ray missions (e.g., \textsl{eXTP}, \textsl{Athena}, and \textsl{HEX-P}) promises to provide unprecedented high-quality data, which will necessary require more sophisticated synthetic reflection spectra than those available today. Our current efforts are thus focused on the development of more advanced reflection spectra for the analysis of observations from the next generation of X-ray missions. The analysis of real X-ray data requires that the model can generate quickly many accurate spectra in order to scan the parameter space and find the best fit. Since this does not seem to be possible keeping the architecture of current reflection models, we are considering the development of radically different reflection models based on neural networks or surrogate models.

\end{document}